\documentclass[aps,prb,singlecolumn,superscriptaddress, longbibliography, nobibnotes]{revtex4-2}
\usepackage{graphicx}
\usepackage{color}
\usepackage{float}
\usepackage{amsmath}
\usepackage{makecell} 
\usepackage{multirow} 
\usepackage{tabularx}
\usepackage{booktabs} 
\usepackage{siunitx} 

\usepackage[breaklinks]{hyperref}
\hypersetup{
	pdfnewwindow=true,      
    colorlinks=true,       
    linkcolor=blue,          
    citecolor=blue,        
    filecolor=blue,      
    urlcolor=blue           
}

\def\kpar{k_\parallel} 
\begin{document}
\title{Engineering 2D material exciton lineshape with graphene/\textit{h}-BN encapsulation}

\author{Steffi Y. Woo}
\thanks{These two authors contributed equally}
\email{steffi.woo@universite-paris-saclay.fr}
\affiliation{Universit\'e Paris-Saclay, CNRS, Laboratoire de Physique des Solides, 91405, Orsay, France}

\author{Fuhui Shao}
\thanks{These two authors contributed equally}
\affiliation{Universit\'e Paris-Saclay, CNRS, Laboratoire de Physique des Solides, 91405, Orsay, France}
\affiliation{State Key Laboratory for Superlattices and Microstructures, Institute of Semiconductors, Chinese Academy of Sciences, Beijing 100083, China} 
\affiliation{College of Materials Science and Opto-Electronic Technology, University of Chinese Academy of Sciences, Beijing 100083, China}

\author{Ashish Arora}
\affiliation{Institute of Physics and Center of Nanotechnology, University of M\"{u}nster, 48149 M\"{u}nster, Germany}
\affiliation{Indian Institute of Science Education and Research, Dr. Homi Bhabha Road, 411008 Pune, India}

\author{Robert Schneider}
\affiliation{Institute of Physics and Center of Nanotechnology, University of M\"{u}nster, 48149 M\"{u}nster, Germany}

\author{Nianjheng Wu}
\affiliation{Universit\'e Paris-Saclay, CNRS, Laboratoire de Physique des Solides, 91405, Orsay, France}
\affiliation{Universit\'e Paris-Saclay, Institut des Sciences Mol\'eculaires d'Orsay, 91405, Orsay, France}

\author{Andrew J. Mayne}
\affiliation{Universit\'e Paris-Saclay, Institut des Sciences Mol\'eculaires d'Orsay, 91405, Orsay, France}

\author{Ching-Hwa Ho}
\affiliation{Graduate Institute of Applied Science and Technology, National Taiwan University of Science and Technology, Taipei, Taiwan}

\author{Mauro Och}
\affiliation{Department of Materials, Imperial College London, London SW7 2AZ, UK}

\author{Cecilia Mattevi}
\affiliation{Department of Materials, Imperial College London, London SW7 2AZ, UK}

\author{Antoine Reserbat-Plantey}
\affiliation{ICFO-Institut de Ci\`encies Fot\`oniques, The Barcelona Institute of Science and Technology, 08860 Castelldefels, Spain}
\affiliation{Universit\'e C\^ote d’Azur, CNRS, CRHEA, Valbonne, France}

\author{Alvaro Moreno}
\affiliation{ICFO-Institut de Ci\`encies Fot\`oniques, The Barcelona Institute of Science and Technology, 08860 Castelldefels, Spain}

\author{Hanan Herzig Sheinfux}
\affiliation{ICFO-Institut de Ci\`encies Fot\`oniques, The Barcelona Institute of Science and Technology, 08860 Castelldefels, Spain}

\author{Kenji Watanabe}
\affiliation{Research Center for Electronic and Optical Materials, National Institute for Materials Science, 1-1 Namiki, Tsukuba 305-0044, Japan}
	
\author{Takashi Taniguchi}
\affiliation{Research Center for Materials Nanoarchitectonics, National Institute for Materials Science,  1-1 Namiki, Tsukuba 305-0044, Japan}

\author{Steffen Michaelis de Vasconcellos}
\affiliation{Institute of Physics and Center of Nanotechnology, University of M\"{u}nster, 48149 M\"{u}nster, Germany}

\author{Frank H. L. Koppens}
\affiliation{ICFO-Institut de Ci\`encies Fot\`oniques, The Barcelona Institute of Science and Technology, 08860 Castelldefels, Spain}
\affiliation{ICREA, Barcelona, Spain}

\author{Zhichuan Niu}
\affiliation{State Key Laboratory for Superlattices and Microstructures, Institute of Semiconductors, Chinese Academy of Sciences, Beijing 100083, China} 
\affiliation{College of Materials Science and Opto-Electronic Technology, University of Chinese Academy of Sciences, Beijing 100083, China}

\author{Odile St\'ephan}
\affiliation{Universit\'e Paris-Saclay, CNRS, Laboratoire de Physique des Solides, 91405, Orsay, France}

\author{Mathieu Kociak}
\affiliation{Universit\'e Paris-Saclay, CNRS, Laboratoire de Physique des Solides, 91405, Orsay, France}

\author{F. Javier Garc\'ia de Abajo}
\affiliation{ICFO-Institut de Ci\`encies Fot\`oniques, The Barcelona Institute of Science and Technology, 08860 Castelldefels, Spain}
\affiliation{ICREA-Instituci\'o Catalana de Recerca i Estudis Avan\c{c}ats, Passeig Llu\'{\i}s Companys 23, 08010 Barcelona, Spain}

\author{Rudolf Bratschitsch}
\affiliation{Institute of Physics and Center of Nanotechnology, University of M\"{u}nster, 48149 M\"{u}nster, Germany}

\author{Andrea Kone\v{c}n\'a}
\email{andrea.konecna@vutbr.cz}
\affiliation{Central European Institute of Technology, Brno University of Technology, Brno 612 00, Czech Republic}
\affiliation{Institute of Physical Engineering, Brno University of Technology, Brno 616 69, Czech Republic}

\author{Luiz~H.~G.~Tizei}
\email{luiz.galvao-tizei@universite-paris-saclay.fr}
\affiliation{Universit\'e Paris-Saclay, CNRS, Laboratoire de Physique des Solides, 91405, Orsay, France}

\maketitle

\textbf{Control over the optical properties of atomically thin two-dimensional (2D) layers, including those of transition metal dichalcogenides (TMDs), is needed for future optoelectronic applications. Remarkable advances have been achieved through alloying \cite{Wang2015alloy,Lu2017janus}, chemical \cite{Mouri2013molecules,Kim2016doping,Ho2017doping} and electrical \cite{Mak2013FET,Chernikov2015} doping, and applied strain \cite{He2013strain,Schmidt2016,Harats2020}. However, the integration of TMDs with other 2D materials in van der Waals heterostructures (vdWHs) to tailor novel functionalities remains largely unexplored. Here, the near-field coupling between TMDs and graphene/graphite is used to engineer the exciton lineshape and charge state. Fano-like asymmetric spectral features are produced in WS$_2$, MoSe$_2$ and WSe$_2$ vdWHs combined with graphene, graphite, or jointly with hexagonal boron nitride (\textit{h}-BN) as supporting or encapsulating layers. Furthermore, trion emission is suppressed in \textit{h}-BN encapsulated WSe$_2$/graphene with a neutral exciton redshift (44 meV) and binding energy reduction (30 meV). The response of these systems to electron-beam and light probes is well-described in terms of 2D optical conductivities of the involved materials. Beyond fundamental insights into the interaction of TMD excitons with structured environments, this study opens an unexplored avenue toward shaping the spectral profile of narrow optical modes for application in nanophotonic devices.}

Interlayer near-field coupling in vdWHs plays a determinant role on the performance of such 2D nanostructures. For example, the characteristics of graphene-based field-effect transistors were dramatically improved in devices encapsulated in \textit{h}-BN \cite{Dean2010graphene-transport,Decker2015graphene-charges,Cadiz2017,Wierzbowski2017}, due to the atomic flatness and low trap density of \textit{h}-BN \cite{Cadiz2017,Wierzbowski2017}. 
Indeed, encapsulation in graphite (Gr) or \textit{h}-BN ensures monolayer flatness down to tens of picometers \cite{Meyer2007,Shao2022}, as well as improved interfacial cleanliness and homogeneous dielectric environment \cite{Rhodes2019}. In addition to these passive environment and dielectric disorder effects, electromagnetic coupling to substrates has also been used to modify exciton transition and binding energies \cite{Borghardt2017}. However, by tuning the nature of the substrate, one could envision even better control over the transitions, as commonly done in nanophotonics. A variety of coupling regimes can be obtained in plasmonic systems. For example, strong coupling between two plasmons having similar energies and linewidths produces two new energy-split states \cite{Nordlander2004}. Coupling of a spectrally broad plasmonic resonance to a sharp excitonic or phononic mode induces a Fano-like resonance, with a modified and possibly asymmetric spectral lineshape \cite{Schlather2013}. In the same line, the engineering of Fano-resonant ultrathin optical multilayers leads to new optical properties with applications in, for example, photovoltaics \cite{elkabbash2021}. Given the ability to realize atomic-layer control of vdWHs, detailed monitoring of their optical properties at high spatial and spectral resolution is indispensable. 

In this work, different electromagnetic environments for TMD monolayers have been designed and precisely fabricated to produce a dramatic modification of their optical response. Optical absorption and spatially resolved electron energy-loss spectroscopy (EELS) spectra show asymmetric lineshapes, similar to Fano profiles at energies close to the excitonic transitions in TMD monolayers when encapsulated or supported on graphene or thin graphite ($<$10 nm thickness). EELS measurements were performed at temperatures of T $\approx$ 110~K in which the transition linewidth is not altered compared to the response of the same monolayers encapsulated in \textit{h}-BN. A simple model based on the 2D optical conductivity \cite{GarciadeAbajo2013}, including retardation corrections for the interaction with the electron beam and combining a TMD monolayer with a conductive or dielectric environment, explains the observed lineshapes. This indicates that: i) coupling mainly has an electromagnetic origin, without significant modifications to the TMD electronic structure; and ii) dissipation and charge transfer to the graphene or graphite layers do not modify the exciton line broadening within the 10 meV energy resolution in the EELS measurements.

\begin{figure} [H]
\centering{\includegraphics[width=1\textwidth]{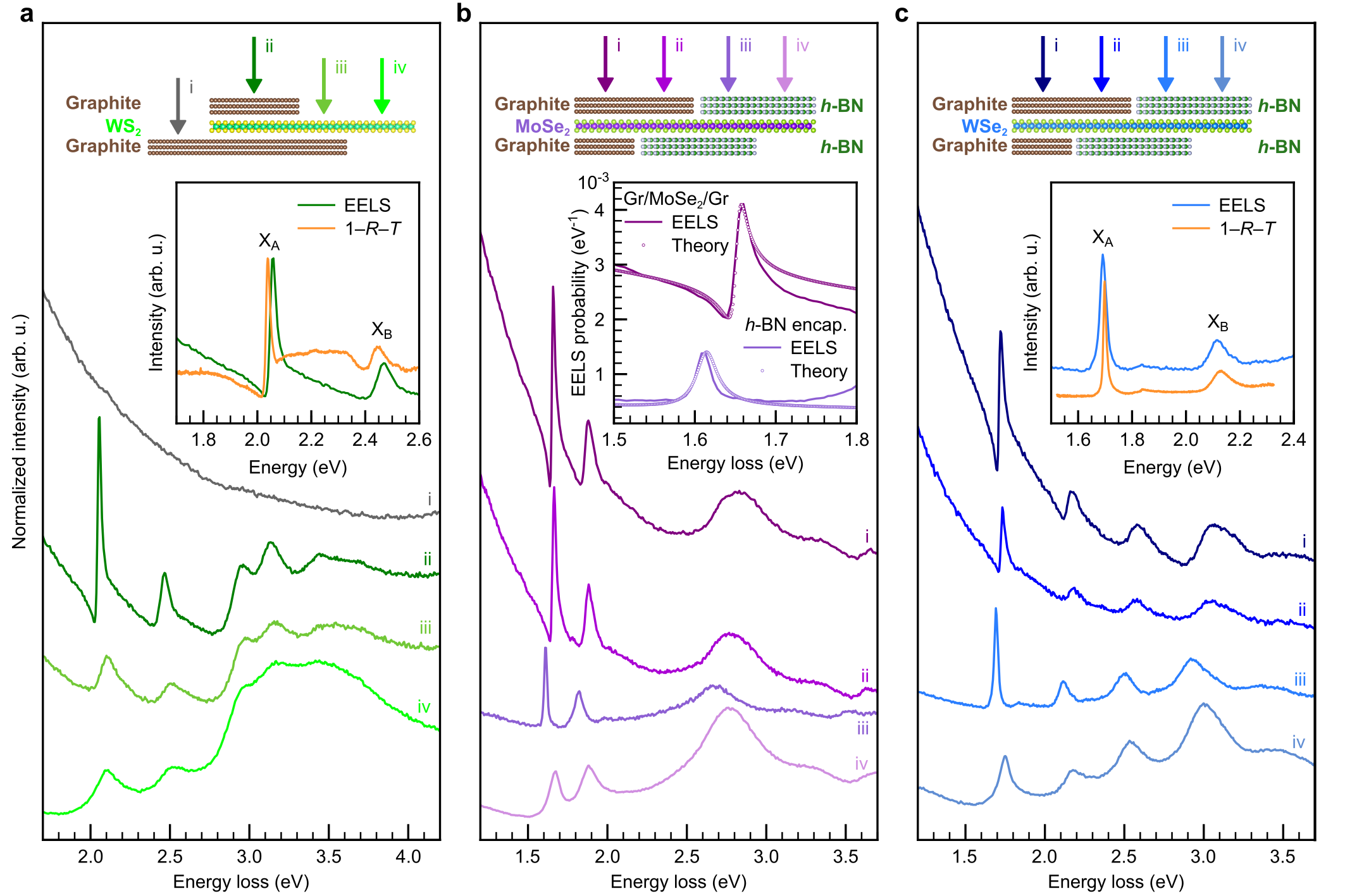}}
\caption{\textbf{Spectra of TMD monolayer with graphite and/or \textit{h}-BN encapsulation}: EELS spectra of (a) WS$_2$, (b) MoSe$_2$, and (c) WSe$_2$ monolayers in different configurations of freestanding, supported or encapsulated with \textit{h}-BN and/or thin graphite measured at $T = 110$~K. The configuration of each spectrum is color-coded by the arrows and roman numerals on the sketches in the upper part of the panels. All spectra are normalized with respect to the total intensity of the elastic (zero-loss) peak (ZLP) and vertically offset for clarity. Insets in (a) and (c) show the optical absorption spectrum (orange curves) on similarly produced Gr/WS$_2$/Gr and \textit{h}-BN/WSe$_2$/\textit{h}-BN heterostructures at 5 K and 150 K, respectively, compared to their respective EELS spectrum. The inset in (b) shows the comparison of experimentally measured (lines) and modelled (dots) EELS spectra for a MoSe$_2$ monolayer encapsulated in thin graphite (asymmetric lineshape) and \textit{h}-BN (Lorentzian lineshape).}
\label{Fig_TMDSpectra}
\end{figure} 

For each of the TMD monolayers considered here (WS$_2$, MoSe$_2$, and WSe$_2$), vdWHs were designed and fabricated with a structure varying along both the in-plane and out-of-plane directions, as represented in the upper sketches of Figs. \ref{Fig_TMDSpectra}(a--c) and \ref{fig_sample-prep} in the Supplementary Information (SI). This ensures that a single TMD monolayer is studied under the same conditions but with varying dielectric environment. The sub-nm sized electron probe used in the EELS experiment is much smaller than the typical lateral extension of different stack configurations, allowing site selectivity when studying them individually and at their interfaces [such as shown in Fig. \ref{fig_temperature}(a)].

Measured EELS and optical absorption spectra of the vdWHs are shown in Fig. \ref{Fig_TMDSpectra} and contain a series of excitonic transitions \cite{Arora2020,Bonnet2021}. The sharpest feature in the spectra is the lowest-energy exciton, named X$_A$, which occurs at the lowest-lying optically active transition at the \(K\)/\(K^{\prime}\) points in reciprocal space. For the analyzed samples, the evolution from freestanding to \textit{h}-BN supported and encapsulated monolayer behaves as reported in the literature \cite{Cadiz2017,Arora2020,Bonnet2021,Shao2022}: the $\sim$100--150 meV broad peak sharpens, attaining a Lorentzian lineshape when encapsulated in \textit{h}-BN [Fig. \ref{Fig_TMDSpectra}(b--c)] with a full-width at half-maximum (FWHM) in the 20--40 meV range at 110 K. Encapsulation between \textit{h}-BN and/or graphite ensures reduced monolayer roughness and fewer adsorbates on TMD monolayers \cite{Shao2022}, as shown for the case of the Gr/WS$_2$/Gr heterostructure in Fig. \ref{fig_roughness} in the SI. 

For TMD monolayers encapsulated in thin graphite (Gr/TMD/Gr), the lineshape is markedly different from those encapsulated in \textit{h}-BN, especially when comparing to spectra from EELS or optical absorption in \textit{h}-BN/WSe$_2$/\textit{h}-BN [Fig. \ref{Fig_TMDSpectra}(c) inset]. Asymmetric lineshapes characteristic of Fano profiles \cite{Fano1961} are observed at slightly redshifted energies compared to the excitonic transitions of the freestanding TMD, as most evident in the two lowest-energy excitons X$_A$ and X$_B$ in the insets. The asymmetric lineshape appears in addition to the known continuous absorption of graphite [Fig. \ref{Fig_TMDSpectra}(a) grey EELS spectrum for bare and thin graphite]. The redshifts of the excitons in Gr/WS$_2$/Gr relative to freestanding WS$_2$ arise from dielectric screening, and in fact 1--2 layered graphene has been used previously to tune the electronic gap and exciton binding energy \cite{Raja2017}. A Gr/WS$_2$/Gr heterostructure supported on a sapphire substrate shows a similar feature at the X$_A$ and X$_B$ exciton energies in the optical absorption spectrum measured at T = 5 K [inset of Fig. \ref{Fig_TMDSpectra}(a)]. Furthermore, the persistence of the asymmetric lineshape at room temperature in EELS for the Gr/WS$_2$/Gr heterostructure [shown in Sec. \ref{sec_temperature} and Fig. \ref{fig_temperature}] suggests a different origin than previously reported, where the continuum stems from the trion state \cite{Arora2015fano}. Coupling occurs for TMD layer thicknesses beyond monolayers as well, leading to similarly asymmetric lineshapes [see results for graphite-encapsulated WS$_2$ bilayer shown in Fig. \ref{Fig_WS2-thickness_MoSe2-Ni}].

The observations so far suggest that the resulting lineshape and its asymmetry depends on the thickness of the encapsulating graphene/graphite layers [see details in Sec. \ref{sec_temperature} and Fig. \ref{WS2-thickness_exp-theory-fitting}(a)]. For a WS$_2$ monolayer supported on graphite, a broad peak (larger than 100 meV FWHM) is observed, similar to that of the suspended monolayer [Fig. \ref{Fig_TMDSpectra}(a)], but with a strongly asymmetric lineshape towards lower energy. The broadened linewidth most likely occurs due to the poor optical response of non-encapsulated layers caused by residues and adsorbates on the remaining free surface \cite{Shao2022}. The consequence of heterostructure interfacial cleanliness on the visibility of the asymmetric Fano lineshape due to linewidth broadening is demonstrated in Gr/TMD/\textit{h}-BN heterostructures of MoSe$_2$ and WSe$_2$ shown in Fig. \ref{Fig_TMDSpectra}(b) and (c), respectively. Full encapsulation ensures locally clean interfaces that can be probed selectively by positioning the electron beam, for which the lineshape asymmetry is also observed, although with a lower contrast for TMDs encapsulated only on one side by thin graphite.
The FWHM of the Fano-like lineshapes, given by the interval between the minimum and maximum of the asymmetric profile, for graphite encapsulation is similar to those with \textit{h}-BN-encapsulation, indicating that the line broadening is not largely influenced by the damping from the graphene/graphite layers.

To summarize the experimental observations in Fig.~\ref{Fig_TMDSpectra}, the lineshapes around the excitonic transitions of X$_A$ and X$_B$ strongly depend on the material and thickness of the encapsulating layers. The observed spectral behavior can be understood by starting with a theoretical description of the optical response of 2D heterostructures within the framework of classical electrodynamics. It is assumed that each $j-$th layer forming the heterostructure is so thin that finite-thickness effects can be neglected. Optical properties of an individual layer are described by a frequency-dependent conductivity $\sigma_j(\omega)=\mathrm{i}\omega t_j/(4\pi)[1-\epsilon_j(\omega)],$ where $t_j$ is the layer thickness, and $\epsilon_{j}$ is its dielectric function depending on optical frequency $\omega$. Stacking the layers together results in the overall optical conductivity of the vdWH modeled as $\sigma(\omega)=\sum_j\sigma_j(\omega)$ \cite{El-Fattah2019}. 

The interaction of electromagnetic waves with the vdWH described by the total optical conductivity is encompassed in the Fresnel coefficients \cite{GarciadeAbajo2014_Graphene_plasmonics}. Incidentally, the Fresnel reflection coefficient for p-polarized light,  $r_\mathrm{p}(\kpar,\omega)=1/[1+\omega/(2\pi\sigma \sqrt{(\omega/v)^2-\kpar^2})]$, where $\kpar$ is the in-plane (with respect to the layer planes) wave vector, also enters the electron energy-loss probability for an electron beam passing through the thin vdWH 
\begin{align}
    \Gamma_{\rm EELS}(\omega)&=\frac{4e^2}{\pi\hbar v^2}\int_0^\infty \frac{\kpar^3\,d\kpar}{\big[\kpar^2+(\omega/v\gamma)^2\big]^2}\; {\rm Re}\left\{\frac{1}{\sqrt{(\omega/v)^2-\kpar^2}}\,r_{\rm p}(\kpar,\omega) \right\},
\label{Eq:Gamma}
\end{align}
where $e$ is the elementary charge, $\hbar$ the reduced Planck constant, $v$ the electron velocity, $c$ is the speed of light in vacuum, and $\gamma=1/\sqrt{1-v^2/c^2}$. Incidentally, a non-retarded expression for the Fresnel coefficient is used as an accurate description for layers of small thickness compared with $c/\omega$. The fully non-retarded expressions can be obtained straightforwardly by setting $c\rightarrow \infty$ and $\gamma= 1$ \cite{GarciadeAbajo2013}.
Although this theoretical description completely neglects microscopic electronic interaction (e.g. orbital hybridization) between the individual atomic planes, such as those captured by first-principle calculations, it can still nicely describe the experimental observations in Fig.~\ref{Fig_TMDSpectra}. Focusing on modelling the spectra around the region of the X$_A$ excitonic transition, for which the Lorentz-Drude model of the optical conductivity is applied: $\sigma_\mathrm{TMD}=\mathrm{i}\omega t_\mathrm{TMD}/(4\pi)[-f/(\omega_\mathrm{A}^2-\omega^2-\mathrm{i}\gamma\omega)]$, where $f$ represents the transition strength, $\omega_\mathrm{A}$ is its angular frequency and $\gamma$ stands for a phenomenological damping. The dielectric response of the encapsulating materials can be approximated by constant values in the considered energy region: $\epsilon_\mathrm{Gr}\approx 6+\mathrm{i}10$, and $\epsilon_{h\textrm{-BN}}\approx 4+\mathrm{i}0.5$ \cite{djurivsic1999optical,lee2019refractive}. The TMD and encapsulating layer thicknesses are further set according to the experimentally estimated values [see Table \ref{Lorentz_fit_parameters} in the SI]. 

The inset in Fig.~\ref{Fig_TMDSpectra}(b) shows the correspondence between experimentally measured spectra and the model for MoSe$_2$, where values of $f$, $\omega_\mathrm{A}$ and $\gamma$ are fitted to take into account small energy shifts and differences in broadening originating from microscopic effects. The fitted values (summarized in Table \ref{Lorentz_fit_parameters}), however, show only small differences for the three modelled spectra, which confirms the validity of this approach. The model reproduces the symmetric Lorentzian peak observed for MoSe$_2$ encapsulated in \textit{h}-BN, while it follows the asymmetric lineshape for the graphite encapsulation. A comparison of all experimentally measured spectra with their modelled counterparts is summarized in Figs. \ref{EELS-prob_exp-theory} and \ref{WS2-thickness_exp-theory-fitting}(a) in the SI. 

\begin{figure}
\centering{\includegraphics[width=\textwidth]{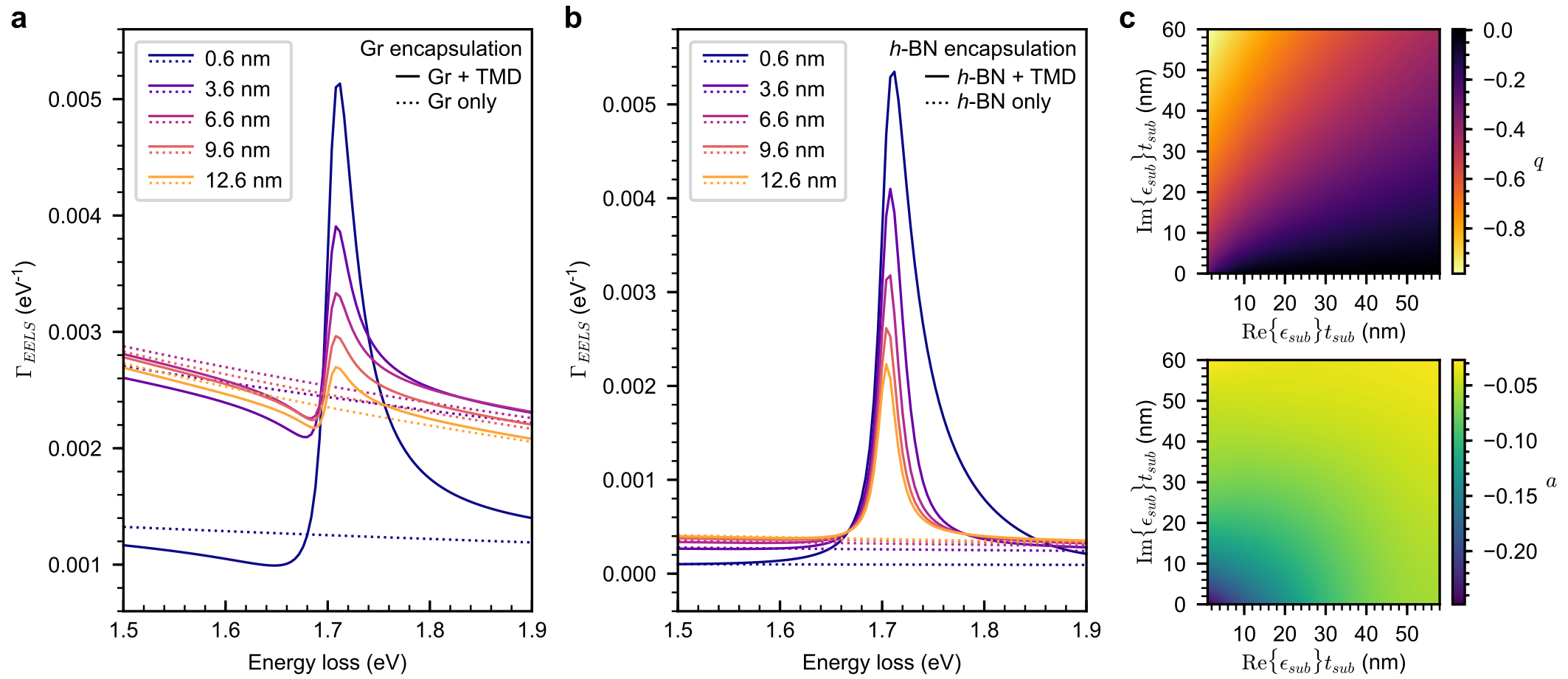}}
\caption{\textbf{Modelled spectra versus parameters of the encapsulating layers}: Evolution of the 2D-modelled EELS spectral shape as a function of (a) graphite and (b) \textit{h}-BN encapsulation thickness. Solid lines represent TMD monolayer + encapsulation, while dashed lines are calculated for encapsulation material only. \textbf{(c)} Fitted parameters \textit{q} and \textit{a} of the modified Fano-like profile [Eq.~\eqref{Eq:Fano_model}] as a function of the encapsulation properties, specifically represented by the real versus imaginary part of the dielectric function multiplied by the thickness of the corresponding encapsulating layer, $\mathrm{Re}\{\epsilon_\textrm{sub}\}t_\textrm{sub}$ and $\mathrm{Im}\{\epsilon_\textrm{sub}\}t_\textrm{sub}$, respectively.}
\label{ModelSpectra}
\end{figure}

The main reason for the asymmetry of the lineshape for graphite encapsulation \textit{versus} the symmetric spectral features with \textit{h}-BN encapsulation stems from the different dielectric response of the two materials. In contrast to $\textit{h}$-BN, which in the spectral region of interest exhibits only negligible values of $\mathrm{Im}[\epsilon_{h\textrm{-BN}}]$, graphite is a quite lossy material. Fig.~\ref{ModelSpectra}(a,b) further elaborates on the lineshape of the EELS probability as a function of graphite and \textit{h}-BN thickness, respectively. By increasing the thickness of the graphite substrate (or encapsulation) attached to a TMD monolayer ($t_\mathrm{TMD}=0.6$~nm) featuring a prototypical excitonic transition characterized by $\hbar\omega_\mathrm{TMD}=1.7$~eV, $\hbar\gamma=0.02$~eV and $f=2$~eV$^2$, there is an increase in the asymmetry (and a decrease in contrast) with respect to a slowly varying background that gradually converges to a smooth spectrum of a freestanding graphite layer (dashed lines). On the contrary, EELS lineshapes for the same TMD layer in the vicinity of \textit{h}-BN are nearly symmetric for all \textit{h}-BN thicknesses.

To obtain more systematic insight into the dependence of the spectral asymmetry on the optical properties of the layers surrounding the TMD monolayer, EELS spectra are calculated for a large range of constant values of $\mathrm{Re}\{\epsilon_\textrm{sub}\}t_\textrm{sub}$ and $\mathrm{Im}\{\epsilon_\textrm{sub}\}t_\textrm{sub}$ representing the real and imaginary parts of the dielectric function of the substrate multiplied by its thickness. The resulting spectra can then be fitted to an empirical model function  
\begin{align}
    \mathcal{F}(\omega)=\frac{1}{\omega}\left[b+a\frac{(q+\Omega)^2}{1+\Omega^2}\right],
    \label{Eq:Fano_model}
\end{align}
where $\Omega=(\omega-\omega_\textrm{TMD})/\gamma$, $b$ represents a constant background, the parameter $a$ is related to the contrast and $q$ to the asymmetry (one recognizes the original Fano lineshape \cite{Fano1961} in the fraction). Fig.~\ref{ModelSpectra}(c) corroborates that the largest asymmetries indeed emerge for large values of $\mathrm{Im}\{\epsilon_\textrm{sub}\}t_\textrm{sub}$. The substrate damping thus clearly opens a coupling channel resulting in the Fano-like asymmetric spectral profile. The increase in asymmetry is however accompanied by the decrease in the contrast as confirmed in Fig.~\ref{ModelSpectra}(c) [see Fig.~\ref{Fig:Fano_fit} in the SI for all parameters extracted from the fitting].

\begin{figure}
    \centering
    \includegraphics[width=12.5cm]{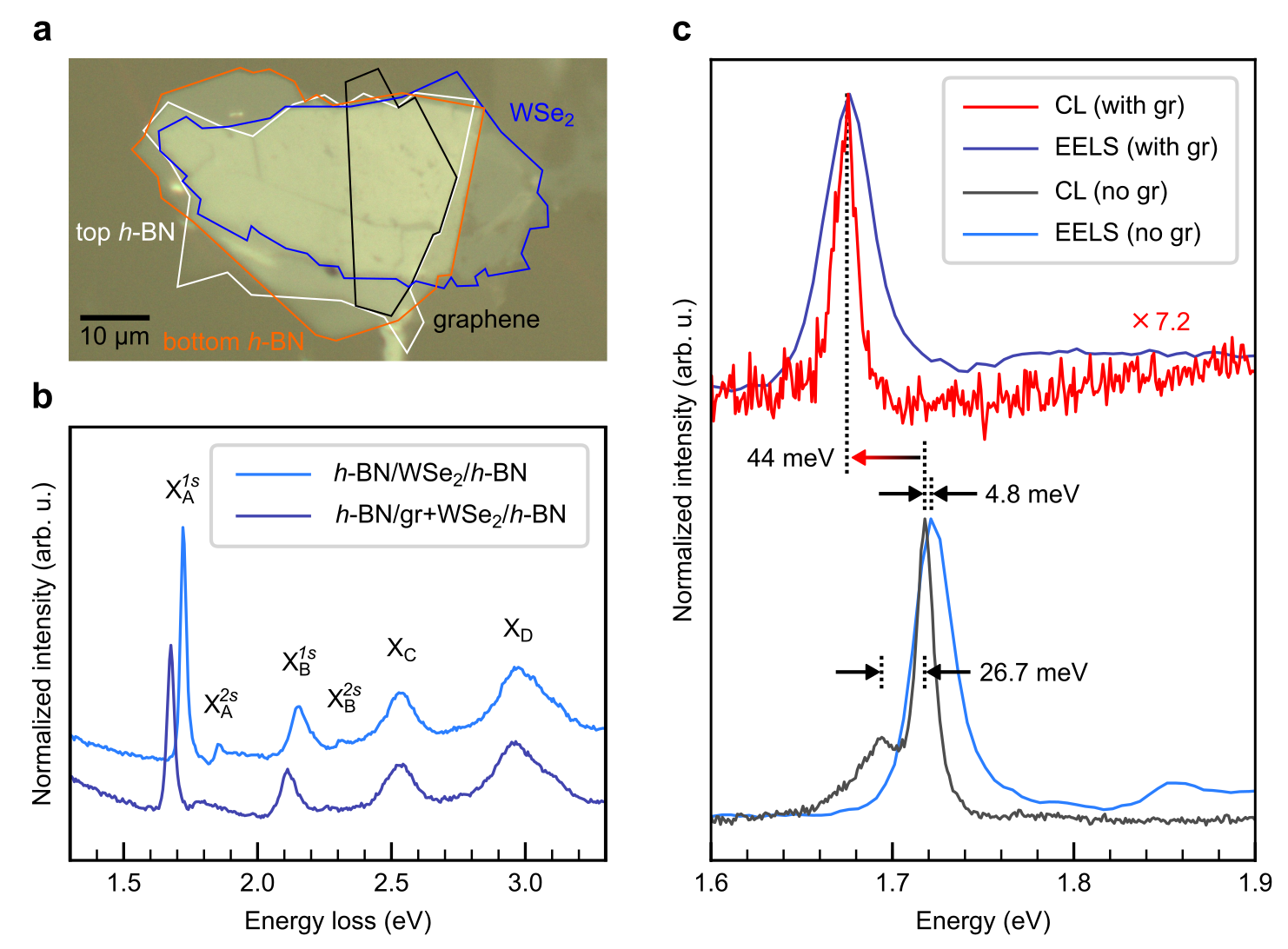}
    \caption{\textbf{Near-field coupling of \textit{h}-BN-encapsulated TMD monolayers with graphene:} (a) Optical micrograph of a \textit{h}-BN/graphene+WSe$_2$/\textit{h}-BN heterostructure with the different constituents outlined. (b) EELS spectra of \textit{h}-BN encapsulated WSe$_2$ and graphene/WSe$_2$ heterostructures, highlighting the clear redshift in the presence of graphene (gr). (c) CL and EELS spectra from identical nanometric regions allows for a clear assignment of the emission peaks, including a prominent low-energy trion emission in the absence of graphene, as well as a clear quenching of the neutral exciton ($\sim$7$\times$ lower) but no obvious trion emission.}
    \label{Fig_CL-EELS_graphene}
\end{figure}

Different vdWH configurations involving graphene or other metals with \textit{h}-BN-encapsulated TMDs were also investigated, where the interlayer distance between the TMD and graphene/metal can be varied by using \textit{h}-BN as a spacer [see Fig. \ref{Fig_CL-EELS_graphene} and Fig. \ref{Fig_WS2-thickness_MoSe2-Ni}(b)]. EELS spectra comparing graphene/WSe$_2$ heterostructure and the same WSe$_2$ monolayer only [both \textit{h}-BN-encapsulated, as seen in the sample optical micrograph of Fig. \ref{Fig_CL-EELS_graphene}(a)] in Fig. \ref{Fig_CL-EELS_graphene}(b) show X$_A$ linewidths spanning the same range (20--40 meV FWHM) and a redshift (47 meV) in the X$_{A}^{1s}$ resonance in the presence of graphene. The first excited states (\textit{n} = 2) of the two lowest-energy ground-state excitons (\textit{n} = 1), namely X$_{A}^{2s}$ and X$_{B}^{2s}$, are also prominently visible in contrast to the graphite-encapsulated WSe$_2$ in Fig. \ref{Fig_TMDSpectra}(c). The energy separation ($\Delta_{12}$) between the $1s$ and $2s$ states of an exciton within its Rydberg series is proportional to the exciton binding energy \cite{Chernikov2014rydberg,Raja2017}, and can be used to approximate the electronic gap. The energy separation for the \textit{h}-BN-encapsulated WSe$_2$ measures $\Delta_{12}$ = 135 meV, whereas it is reduced to $\Delta_{12}$ = 120 meV for the graphene/WSe$_2$ heterostructure, indicative of a reduction in exciton binding energy by 20--30 meV for the latter if considering $E_B \approx 1.3\Delta_{12}-2\Delta_{12}$ for the non-hydrogenic Rydberg series for WSe$_2$ \cite{He2014rydberg,Stier2018rydberg-bindingE}.

The integration of graphene with \textit{h}-BN encapsulation also allows for the opportunity to carry out combined electron spectroscopies of EELS plus cathodoluminescence (CL) on identical regions with subwavelength resolution. The CL spectrum from \textit{h}-BN-encapsulated WSe$_2$ monolayer region is shown in Fig. \ref{Fig_CL-EELS_graphene}(c, bottom). It contains two emission lines that are assigned to the neutral exciton X$_A^0$ (1.718 eV) and trion X$^*$ (1.691 eV), respectively. The Stokes shift of 4.8 meV for the WSe$_2$ monolayer is consistent with optical measurements \cite{Niehues2020}. Severe quenching in the CL signal was observed in graphene/WSe$_2$ [Fig. \ref{Fig_CL-EELS_graphene}(c, top)], but the remaining single emission line, which is redshifted by 44 meV, can be rightfully assigned to X$_A^0$ when compared to the EELS absorption with no obvious low-energy X$^*$ shoulder. This observation is in agreement with low-temperature photoluminescence (PL) from TMD/graphene heterostructures that has shown such a single emission line from only X$_A^0$ for MoS$_2$, WS$_2$, MoSe$_2$, and WSe$_2$ \cite{Lorchat2020}.

The neutralization effect was attributed to the competition of several relaxation pathways and their lifetime differences, between the neutral (2.3 ps for MoSe$_2$) and charged exciton (30 ps for MoSe$_2$) relative to the non-radiative transfer to the graphene \cite{Loan2014,Lorchat2020}. In addition to this, the binding energy of the neutral exciton is expected to decrease in the presence of graphene due to additional Coulomb screening \cite{Raja2017}, which increases the radiative lifetime \cite{Lorchat2020}. 
For heterostructures with the TMD monolayers in contact with nm-thick graphite, no CL emission was observed, contrary to structures incorporating \textit{h}-BN-encapsulation \cite{Zheng2017, Bonnet2021}. Altogether, the comparison of the effect produced by graphene versus graphite can be rationalized in the following way: in the case of graphite, a larger non-radiative damping is at the origin of both the emergence of Fano-like lineshapes and the disappearance of radiative emission; with the influence of the graphene being weaker, radiative emission is still possible but weaker and charge-state selective, and logically the Lorentzian lineshape is preserved. 

Finally, it is emphasized that the here-described Fano effect is different from previously observed Fano effects in the context of TMD spectra. In contrast to previously reported Fano-like lineshapes in TMDs \cite{Arora2015fano}, the continuum is extrinsic to the TMDs, allowing full control on the effect. As compared to Fano-like coupling in plasmonic systems, not only are the continuum and discrete state more rigorously defined, but also the (non-)radiative nature of these two states are reversed between TMD vdWHs and plasmons in nanoparticles.

In summary, the near-field coupling between mono- and few-layered TMDs and graphene/graphite with/without \textit{h}-BN was experimentally shown to enable a customization of the TMD exciton lineshapes, in excellent agreement with theory. Asymmetric exciton lineshapes along with narrow linewidths were reported for vdWHs with thin graphite or graphene encapsulation using both cryogenic optical absorption and EELS. Such a response can be explained through a simple 2D optical conductivity model of the heterostructure.
Interfacing graphene (or other metals) in combination with \textit{h}-BN encapsulation under different vdWH configurations offers additional flexibility in the control of interlayer separation. This further enabled complementary electron spectroscopies of EELS and CL, wherein for WSe$_2$ monolayer in the presence of graphene, suppressed charged exciton CL emission and a reduction in exciton binding energy (and hence the quasi-particle gap) of a few tens of meV were observed.
Understanding and exploiting the coupling between TMDs and graphene enables the engineering of the exciton lineshape, as well as the electronic bandgap and exciton binding energy, thus opening the agenda for the development of disruptive excitonic applications..

\section*{Acknowledgement}

This project has been funded in part by the National Agency for Research under the program of future investment TEMPOS-CHROMATEM (reference no. ANR-10-EQPX-50) and the JCJC grant SpinE (reference no. ANR-20-CE42-0020). This project has received funding from thee European Union’s Horizon 2020 research and innovation programme under grant agreement No. 823717 (ESTEEM3) and 101017720 (EBEAM). A.K. acknowledges the support of the Czech Science Foundation GACR under the Junior Star grant No. 23-05119M. A.A. acknowledges financial support from the German Research Foundation (DFG Projects No. AR 1128/1-1 and No. AR 1128/1-2), NM-ICPS of the DST (Government of India) through the I-HUB Quantum Technology Foundation (Pune, India), Project No. CRG/2022/007008 of SERB (Government of India), and MoE-STARS project No. MoE-STARS/STARS-2/2023-0912 (Government of India). N.W. thanks the MAGMA project for funding (ANR-16-MAGMA-0027). C.M. acknowledges the award of a Royal Society University Research Fellowship (UF160539) and the Research Fellow Enhancement Award 2017 (RGF - 180090) by the Royal Society UK. K.W. and T.T. acknowledge support from the JSPS KAKENHI (Grant Numbers 21H05233 and 23H02052) and World Premier International Research Center Initiative (WPI), MEXT, Japan. F.H.L.K, A.M. and A.R.-P. acknowledge BIST Ignite Programme grant from the Barcelona Institute of Science and Technology (QEE2DUP). F.J.G.A. acknowledges support from the European Research Council (Advanced Grant No. 789104-eNANO) and the Spanish MICINN (PID2020–112625 GB-I00 and Severo Ochoa CEX2019-000910-S).

\bibliography{Fano_biblio.bib}

\newpage
	\renewcommand\thefigure{SI\arabic{figure}}
 	\renewcommand{\thesubsection}{S\arabic{subsection}}
	\setcounter{figure}{0} 
	\setcounter{section}{0}
	\section*{Supplementary Information}

\subsection{Methods}
\subsubsection{Sample Preparation}
\begin{figure}[H]
    \centering
    \includegraphics[width=14.5cm]{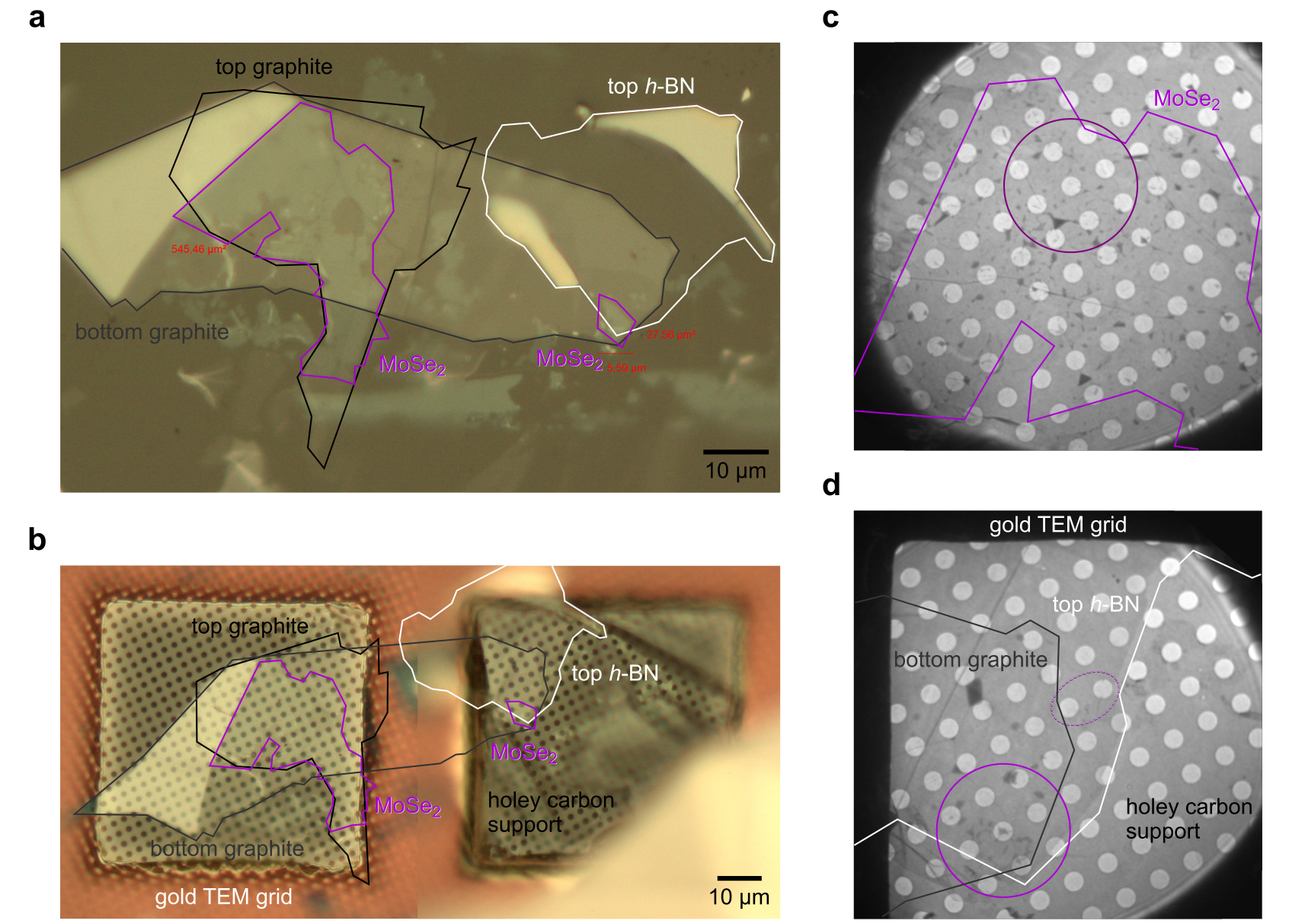}
    \caption{\textbf{Fabrication of the mixed encapsulation sample of MoSe$_2$ monolayer.} (a) Optical micrograph of a \textit{h}-BN/ graphite + MoSe$_2$ + graphite heterostructure imaged on the polymer stamp with the different layer constituents outlined. (b) Optical micrograph of the same heterostructure after dropping onto a holey carbon support TEM grid and washing off any polymer residues. Micrographs in (a,b) are each spliced from two separate images due to differences in focus in the top graphite and top \textit{h}-BN areas. (c) Bright-field STEM image of the heterostructure with the Gr/MoSe$_2$/Gr measurement regions circled. (d) Bright-field STEM image of the \textit{h}-BN/MoSe$_2$/Gr region of the vdWH stack, with the \textit{h}-BN/MoSe$_2$/Gr and \textit{h}-BN/MoSe$_2$ measurement areas circled in solid and dotted lines, respectively. Holes in the TEM grid carbon support are 1.2 $\mu$m in size. Folds, trapped dirt and bubbles of few hundreds of nm size can be observed in the images of (c,d) as dark contrast.}
    \label{fig_sample-prep}
\end{figure}

The samples were prepared using a (bisphenol A polycarbonate) polymer-assisted viscoelastic stamping method to pick up the individual layers making up each vdWH, then dropped onto TEM holey carbon support grids or sapphire substrates directly from the polymer stamp \cite{Zomer2014}. An example of the fabrication and overview of a sample of a single TMD monolayer with mixed encapsulation material at select steps in the procedure is shown in Fig. \ref{fig_sample-prep}. Large-area MoSe$_2$ and WSe$_2$ monolayers were exfoliated from bulk crystals following a gold-mediated exfoliation method onto SiO$_2$/Si substrates \cite{Desai2016}, while WS$_2$ monolayers were grown by chemical vapor deposition. Encapsulating graphite/graphene and \textit{h}-BN flakes were produced from conventional tape exfoliation of bulk crystals sourced from NGS Naturgraphit GmbH and synthesized by the high-pressure high-temperature method \cite{Taniguchi2007}, respectively.

\subsubsection{Electron Spectroscopy and Electron Diffraction}
The monochromated EELS, CL, and electron diffraction were performed on a modified Nion HERMES-S200 (also known as ChromaTEM) operated at 60 keV with the sample at liquid nitrogen temperatures (T $\approx$ 110 K) unless otherwise indicated at room temperature (RT). For EELS and CL experiments, 10 -- 15 mrad convergence angle was used and 1 mrad for nano-diffraction experiments. The EELS spectra were recorded onto a Quantum Detectors MerlinEELS Medipix3 direct electron detector with 256 $\times$ 256 pixel-sized chips in a 4 by 1 geometry. The CL was collected using an Attolight M\"{o}nch system fitted with a 150 groves/mm diffraction grating to give a wavelength resolution of 0.34 nm ($\sim$0.8 meV at 1.72 eV or 720 nm wavelength), recorded onto a Princeton Instruments ProEM EMCCD camera. Different stacking configurations in measured regions of hundreds of nm$^2$ were determined roughly from the optical micrographs from sample preparation and confirmed locally with nm-specificity by a combination of electron diffraction and core-loss EELS. Graphite and \textit{h}-BN encapsulation layer thicknesses [$t_\mathrm{sub}$] were determined experimentally using EELS log-ratio method \cite{Egerton2011} and calculated values for their inelastic mean free path \cite{Malis1988} with the input parameters (incident electron energy, EELS detector collection semi-angles) set according to the experimental conditions while accounting for contributions from the TMD monolayer within the measured effective mean free path. 

\subsection{Monolayer Roughness with Graphite Encapsulation}
\begin{figure}[H]
    \centering
    \includegraphics[width=0.7\textwidth]{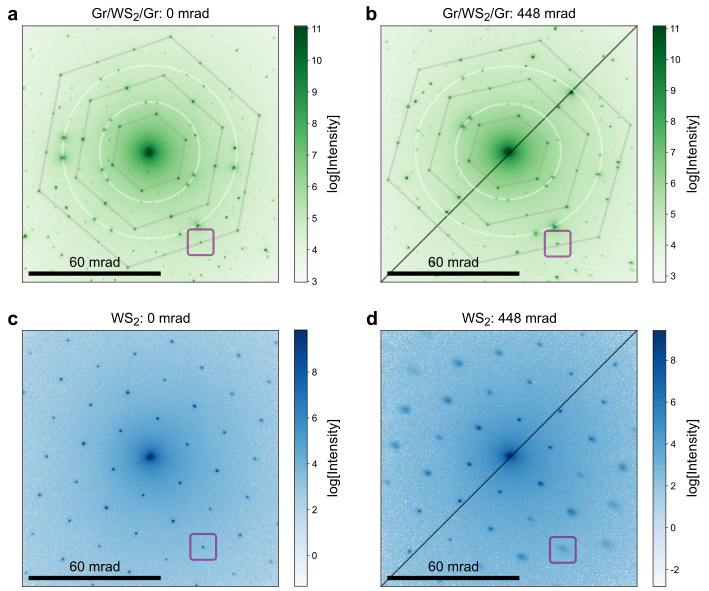}
    \caption{Nano-beam electron diffraction patterns of WS$_2$ monolayer in (a,b) Gr/WS$_2$/Gr and (c,d) freestanding WS$_2$ configurations at sample tilt-angles of 0 and 448 mrad. The diffraction spots boxed in purple are of the same index order used to compare the roughness of the WS$_2$ monolayer. Diffraction spots highlighted by grey hexagons and white circles are from WS$_2$ and graphite, respectively.}
    \label{fig_roughness}
\end{figure}

Nano-beam electron diffraction spots at high sample tilt-angle show no discernible broadening in the Gr/WS$_2$/Gr heterostructure [Fig. \ref{fig_roughness}(b)] as compared to the freestanding WS$_2$ monolayer [Fig. \ref{fig_roughness}(d)]. This evidences that thin graphite layers have comparable capabilities to reduce corrugation in atomically-thin layers as \textit{h-}BN layers as encapsulation material \cite{Shao2022}.

\subsection{Optical Spectroscopy}
\label{sec_optical}
\begin{figure}[H]
    \centering
    \includegraphics[width=7.98cm]{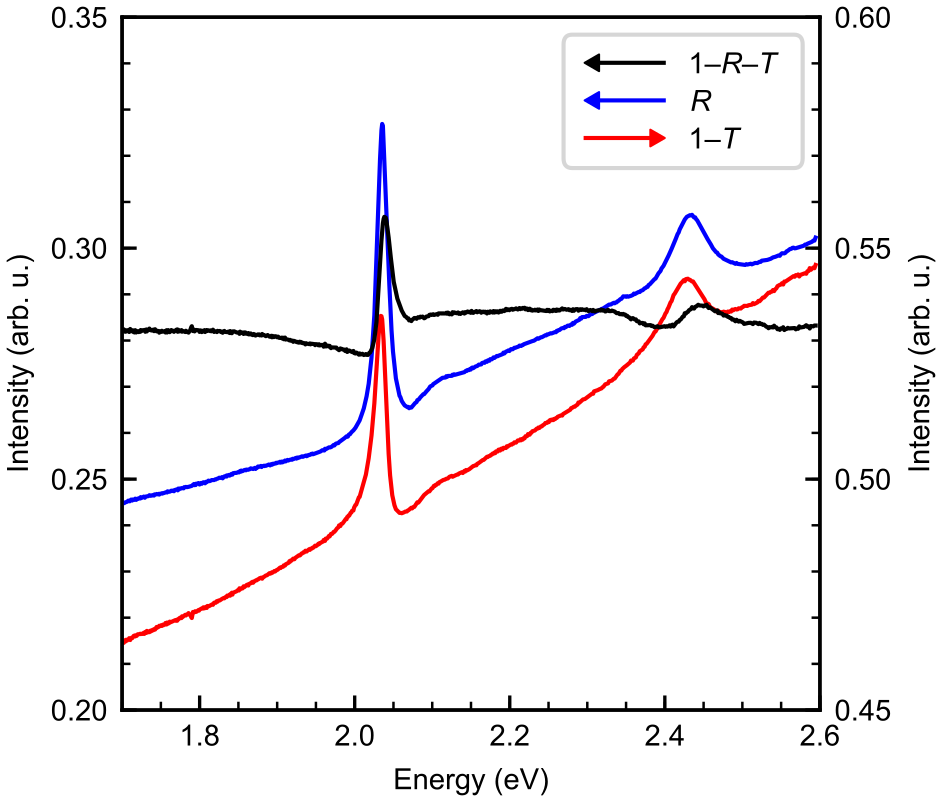}
    \caption{Optical spectroscopy of graphite-encapsulated WS$_2$ monolayer on sapphire substrate measured at 5 K, highlighting the evident asymmetric lineshape in the absorption spectrum ($1 - R - T$) in black, and less evident in both the reflectance ($R$) and transmittance ($1 - T$) spectra in blue and red, respectively.}
    \label{fig_opt-spectra}
\end{figure}

To retrieve the absorption spectrum $A$($\lambda$), reflectance $R$($\lambda$) and transmittance $T$($\lambda$) spectra are measured on the same location of the graphite or \textit{h}-BN encapsulated TMD sample transferred onto sapphire substrates under the same experimental conditions. The optical absorption of the sample is thus calculated as $A(\lambda) = 1 - R(\lambda) - T(\lambda)$. The advantage is that the optical absorption spectral lineshapes obtained by this method can be analyzed directly without taking into account interference effects that arise from multiple interfaces in such samples \cite{Arora2019trions}. These interference effects remain visible within the $R(\lambda$) and  $T(\lambda$) spectra, as shown in the low-temperature (T = 5 K) comparison alongside the $A(\lambda$) spectrum in Fig. \ref{fig_opt-spectra}. Most notably, any asymmetric lineshape is not apparent in the $R(\lambda$) and  $T(\lambda$) spectra, however, the asymmetry is unambiguously present in the $A(\lambda)$ spectrum on the X$_A$ peak.

\subsection{Effect of Temperature}
\label{sec_temperature}
Fano-type resonance at the X$_A$ transition in MoSe$_2$ monolayers with a negative asymmetry parameter (dip at higher energy) by reflectance contrast at T = 5 K has been previously reported \cite{Arora2015fano}. The progressive transition towards a Lorentzian lineshape and disappearance of the charged exciton (trion, X$^*$) peak with increasing temperature proposes the unusual lineshape is caused by the interaction of the ground-state $1s$ X$_A^0$ with the quasi-continuum of the trion X$^*$ excited states. The binding energy of trions are of the order of $\sim$35 meV for WS$_2$, and therefore diminishes in oscillator strength towards room-temperature \cite{Arora2019trions}.

\begin{figure}[H]
    \centering
    \includegraphics[width=\textwidth]{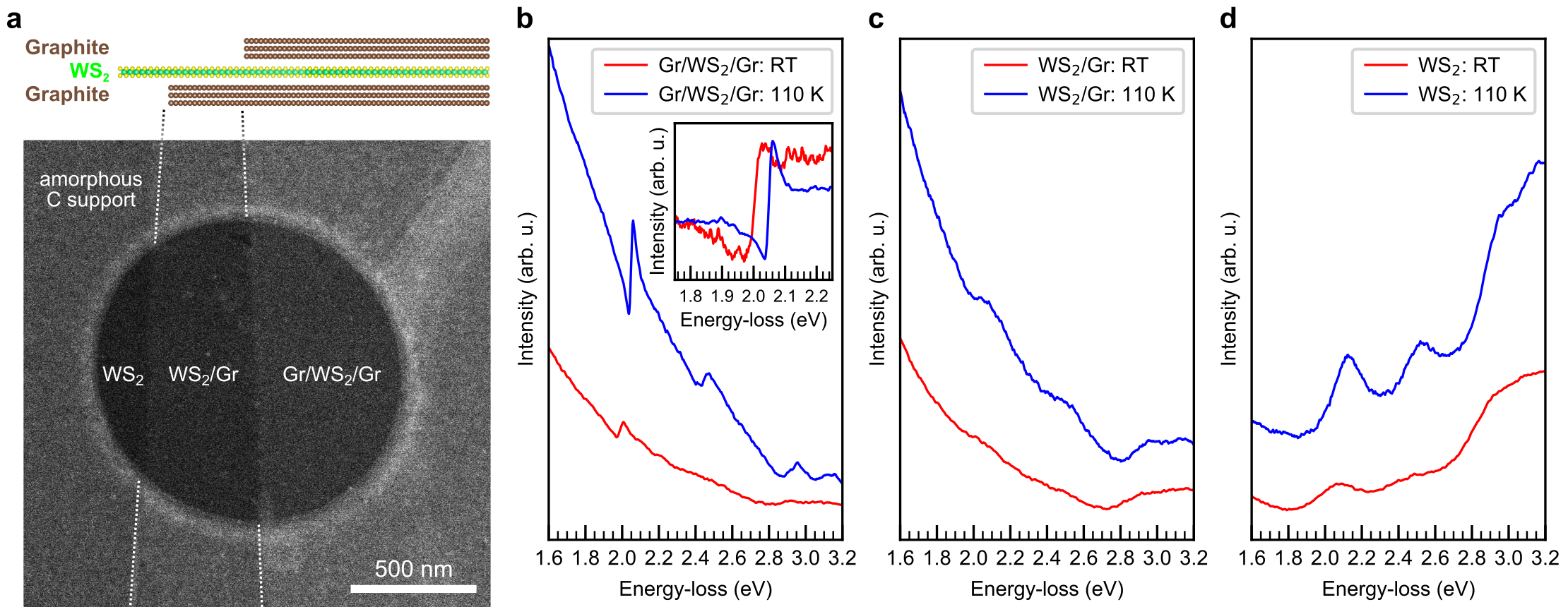}
    \caption{An area including WS$_2$, WS$_2$/Gr, Gr/WS$_2$/Gr measured at two different temperatures. (a) annular dark-field (ADF) image of the region with its partition on top; (b) the spectra of Gr/WS$_2$/Gr, the inset is background subtracted by the Lorentzian tail of graphite; (c) The spectra of WS$_2$/Gr; (d) the spectra of freestanding WS$_2$ monolayer.}
    \label{fig_temperature}
\end{figure}

Room-temperature EELS measurement on identical areas of the Gr/WS$_2$/Gr heterostructure from Fig. \ref{Fig_TMDSpectra}(a) continues to exhibit the asymmetric Fano-lineshape as presented in Fig. \ref{fig_temperature}(b), albeit with significantly lower oscillator strength. A comparison of WS$_2$/Gr and freestanding WS$_2$ monolayer between low- and room-temperature also identifies the same decrease in EELS signal intensity, in addition to the expected exciton linewidth broadening, shown in Fig. \ref{fig_temperature}(c) and (d). The persistence of the asymmetric lineshape at room temperature for the Gr/WS$_2$/Gr heterostructure suggests that the observed asymmetric lineshapes do not originate from coupling to trions.

\subsection{Effect of TMD Layer Thickness}
\begin{figure}[H]
    \centering
    \includegraphics[width=12.9cm]{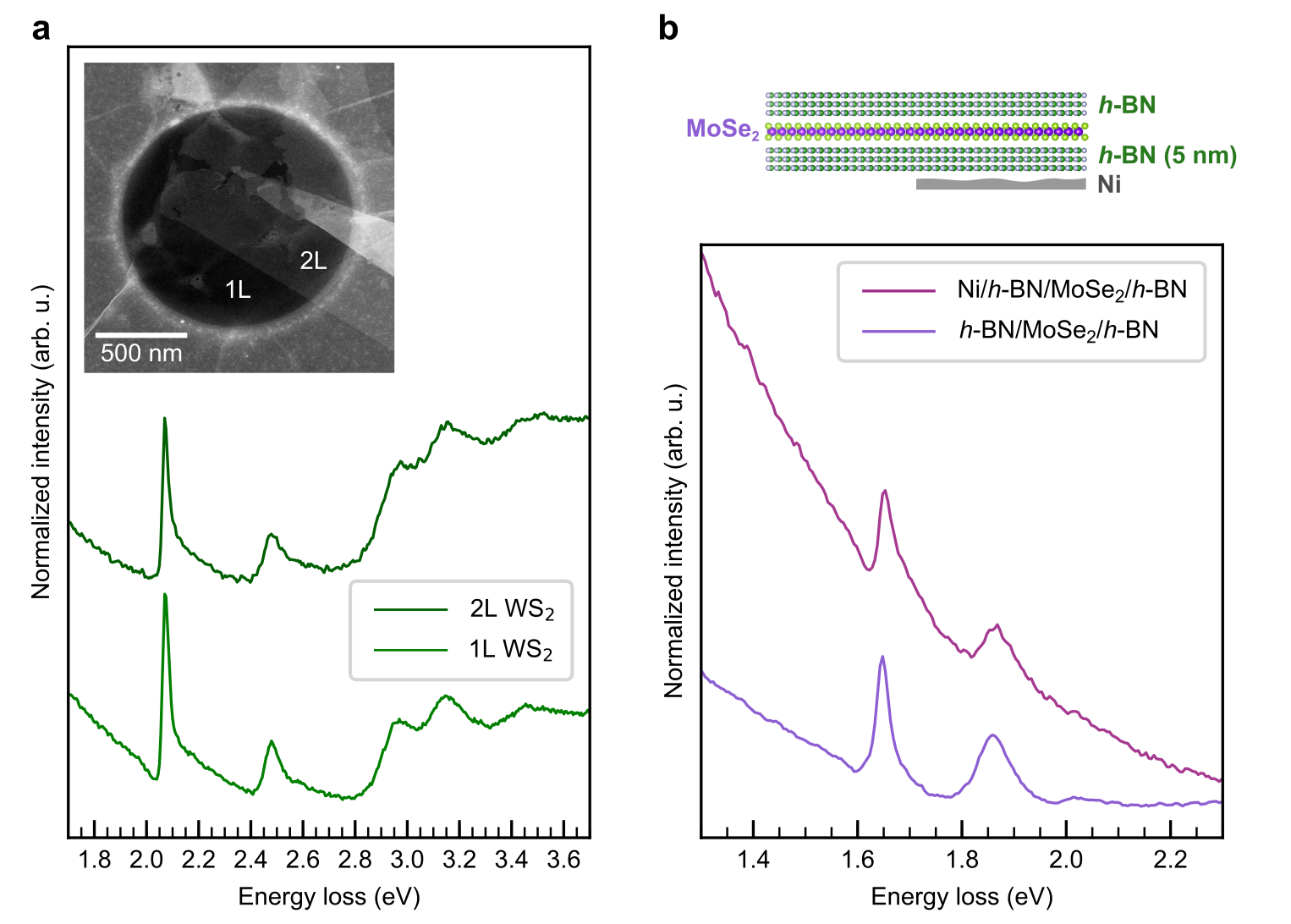}
    \caption{\textbf{Near-field coupling of graphite-encapsulated TMD few-layers and \textit{h}-BN encapsulated TMD monolayers with other metals.} (a) EELS spectra of monolayer (1L) and bilayer (2L) WS$_2$ encapsulated between graphite flakes, inset with the annular dark-field (ADF) image of the measurement regions marked; (b) EELS spectra of \textit{h}-BN encapsulated MoSe$_2$ monolayer with and without Ni metal on the 5 nm-thick \textit{h}-BN side also showing the same asymmetric lineshape.}
    \label{Fig_WS2-thickness_MoSe2-Ni}
\end{figure}

Discrete excitonic transitions in few-layered TMDs encapsulated in thin graphite also exhibit identical asymmetric lineshapes. For example, the X$_A$ and X$_B$ exciton peaks from the Gr/WS$_2$/Gr heterostructure with 1--2 layered WS$_2$ [shown in Fig. \ref{Fig_WS2-thickness_MoSe2-Ni}(a)] are clearly asymmetric for both WS$_2$ layer thicknesses. The graphite encapsulation thickness is the same at both monolayer (1L) and bilayer (2L) regions, as such, their asymmetric lineshape does not change significantly as a function of TMD layer thickness.

\subsection{Coupling between TMD and Other Metals}
Bulk metallic thin films, such as nickel (Ni), deposited onto a continuous silicon nitride support can be used to mimic typical electrical contacts in optoelectronic devices while allowing variable TMD/metal interlayer separation from the \textit{h}-BN spacer thickness. This configuration is also optimal for maintaining the TMD atomic flatness by the use of the \textit{h}-BN \cite{Shao2022}. \textit{h}-BN encapsulated MoSe$_2$ monolayer was deposited onto a 4 nm-thick Ni thin film, with the bottom \textit{h}-BN spacer of 5 nm. The EELS spectrum in the presence of Ni also exhibits clear asymmetric lineshape in Fig. \ref{Fig_WS2-thickness_MoSe2-Ni}(b).

\subsection{Effect of Graphite Thickness}
\label{sec_thickness}

\begin{table}[H]
\centering
\caption[The dielectric function of TMDs and graphite]{\textbf{Encapsulation layer thicknesses and parameters of the dielectric function of TMDs and graphite for various vdWH measured.} Total encapsulation layer thicknesses [$t_\mathrm{sub}$] are determined experimentally (using EELS log-ratio method \cite{Egerton2011} and calculated values for mean free path \cite{Malis1988} in the case of mixed material encapsulation, both graphite/\textit{h}-BN thickness are listed separately. Values of $\epsilon_\mathrm{graphite}$ are represented by the in-plane response of graphite taken from literature \cite{djurivsic1999optical} and approximated by a constant in a given energy range, the rest of the parameters are found via fitting to the form of a Lorentz oscillator.}
\begin{ruledtabular}
\begin{tabular}{lcccccc}
    vdWH configuration & \makecell{$t_\mathrm{sub}$ (nm)} & \makecell{$\omega_\mathrm{0}$ (eV)} & \makecell{$f_\mathrm{TMD}$ (eV$^{2}$)} & \makecell{$\gamma$ (meV)}& $\epsilon_\mathrm{graphite}$ & \makecell{$t_\mathrm{TMD}$ (nm)} \\
     \hline
    graphite/WS$_2$/graphite & 17.5 & 2.05\ & 1.6\ & 30 & $5.34+8.74$i & 0.6 \\
    graphite/WS$_2$/graphite & 6.1 & 2.05\ & 2.8\ & 15 & $5.34+8.74$i & 0.6 \\    
    graphene/WS$_2$/graphene & 0.69 & 2.06\ & 0.6\ & 30 & $5.34 +8.74$i & 0.6 \\
     \hline
    graphite/MoSe$_2$/graphite & 8.3 & 1.65\, & 3.0\ & 10 & $5.91+10.13$i & 0.6 \\
    graphite/MoSe$_2$/\textit{h}-BN & 4.1, 9.6 & 1.66\ & 3.0\, & 20 & $5.91 +10.13$i & 0.6 \\
    \textit{h}-BN/MoSe$_2$/\textit{h}-BN & 28.6 & 1.61\ & 3.0\, & 30 & - & 0.6 \\
     \hline       
    graphite/WSe$_2$/graphite& 20.2 & 1.71\, & 10\ & 20 & $5.84+9.92$i & 0.6 \\
    graphite/WSe$_2$/\textit{h}-BN& 20.9, 45.7 & 1.72\ & 10\ & 20 & $5.84+9.92$i & 0.6\\
    \textit{h}-BN/WSe$_2$/\textit{h}-BN& 32.7 & 1.69\ & 5.7\ & 40 & - & 0.6 \\
\end{tabular}
\end{ruledtabular}
\label{Lorentz_fit_parameters}
\end{table}

\begin{figure}[ht]
    \centering
    \includegraphics[width=12.85cm]{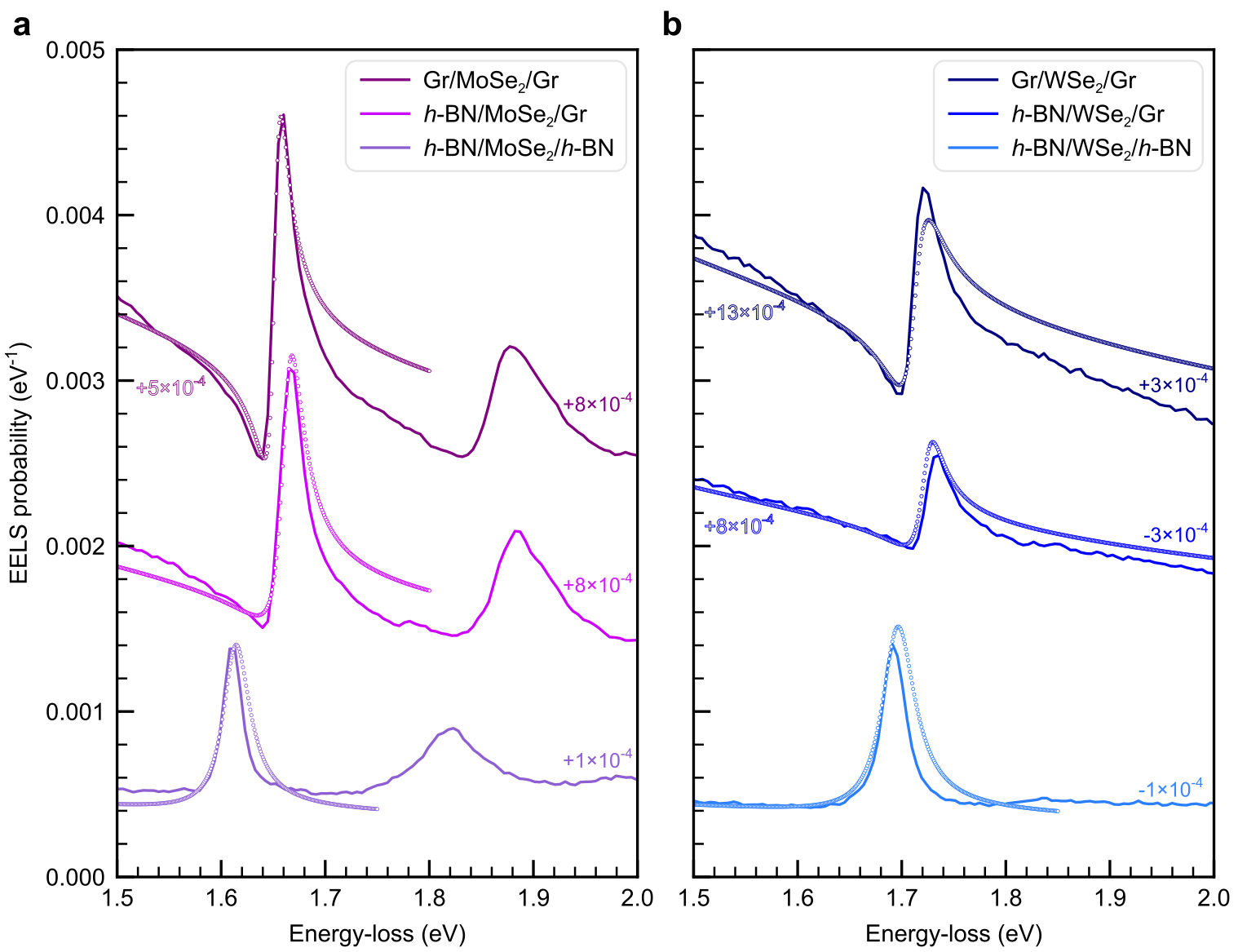}
    \caption{\textbf{Comparison of experimental EELS spectra with the theoretical model.} EELS spectra of mixed graphite and/or \textit{h}-BN encapsulation of (a) MoSe$_2$ and (b) WSe$_2$ monolayers from experiments (solid lines) and the 2D optical conductivity model (dots) obtained by using Eq.~\eqref{Eq:Gamma}. The experimental data is plotted as EELS probability $\Gamma_{\rm EELS}(\omega_{i})=I(\omega_{i})/\int_0^{\omega_{i}} I(\omega)d\omega$, where $\mathcal{I(\omega)}$ is the measured EELS intensity at energy $\omega$. All experimental spectra are reproduced from Fig. \ref{Fig_TMDSpectra}(b) and (c) following the same color-code. Vertical offsets for selected experimental and modelled curves have been added for clarity, and are noted adjacent their curves in solid and open color-coded text, respectively.}
\label{EELS-prob_exp-theory}
\end{figure}

\begin{figure}[ht]
    \centering
    \includegraphics[width=12.85cm]{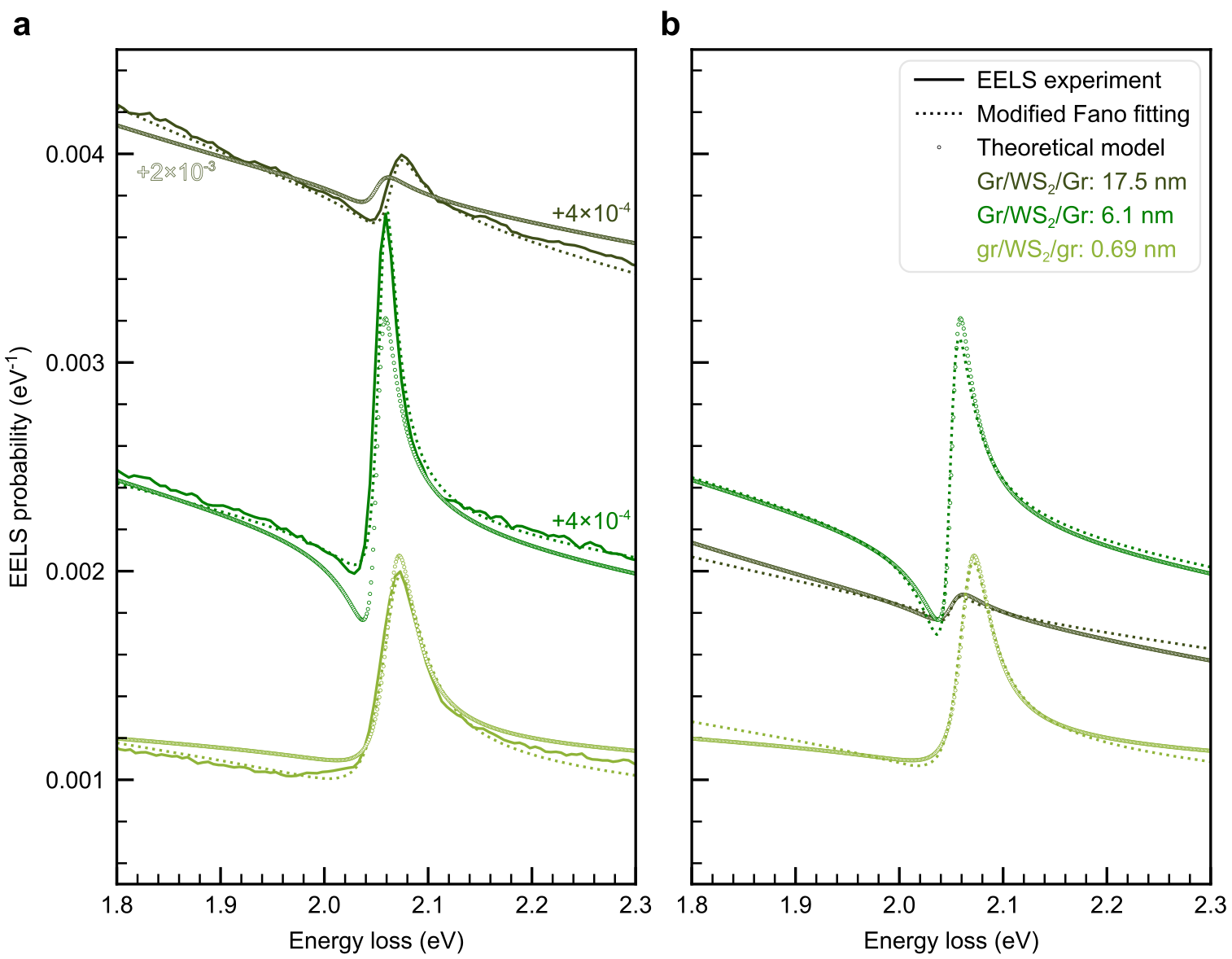}
    \caption{\textbf{Comparison between experimental and modelled EELS spectra of monolayer WS$_2$ of different graphite thickness with their respective fitting with the modified Fano function.} (a) EELS spectra of WS$_2$ monolayer encapsulated between two graphite layers of total thickness of 17.5 nm (thick), 6.1 nm (thin), and two graphene layers (0.69 nm) from experiments (solid lines), the 2D optical conductivity model (dots) obtained using Eq.~\eqref{Eq:Gamma}, and fitting of the experimental data with the modified Fano-like function (dotted lines) in Eq.~\eqref{Eq:Fano_model}. (b) Modelled EELS spectra (dots, reproduced from (a)) obtained using Eq.~\eqref{Eq:Gamma} and its fitting with the modified Fano-like function given in Eq.~\eqref{Eq:Fano_model} (dotted lines). The experimental data is plotted as EELS probability $\Gamma_{\rm EELS}(\omega_{i})=I(\omega_{i})/\int_0^{\omega_{i}} I(\omega)d\omega$, where $\mathcal{I(\omega)}$ is the measured EELS intensity at energy $\omega$. The spectrum of the 6.1 nm (thin) Gr-encapsulation thickness is reproduced from Fig. \ref{Fig_TMDSpectra}(a). The fitted parameters used in the modified Fano-like function curves are listed in Table \ref{Fano_fit_parameters}. Vertical offsets for selected modelled curves and experimental/fitted curve pairings have been added for clarity in (a), and are noted adjacent to their curves in open and solid color-coded text, respectively.}
\label{WS2-thickness_exp-theory-fitting}
\end{figure}

\begin{table}[H]
\sisetup{round-mode=places,round-precision=3}
\centering
\caption{\textbf{Graphite thicknesses and fitting parameters of the modified Fano function of graphite-encapsulated WS$_2$.} Total graphite/graphene encapsulation layer thicknesses [$t_\mathrm{sub}$] are determined experimentally (using EELS log-ratio method \cite{Egerton2011} and calculated values for mean free path \cite{Malis1988}, and fitting parameters of the experimental and modelled EELS spectra using the modified Fano-like function from Eq.~\eqref{Eq:Fano_model}.}
\begin{tabularx}{1\linewidth}%
  {p{0.11\linewidth}%
  p{0.075\linewidth}%
  c*{10}{>{\centering\arraybackslash}X}}
    \toprule
    \toprule
    & & Exp. & Theory & Exp. & Theory & Exp. & Theory & Exp. & Theory & Exp. & Theory\\
    \cmidrule{3-4} \cmidrule(l{.75em}r{0.5em}){5-6} \cmidrule(r{.5em}l{.5em}){7-8} \cmidrule(r{.5em}l{.5em}){9-10} \cmidrule(r{.5em}l{.5em}){11-12}
    Configuration & $t_\mathrm{sub}$ (nm) & \multicolumn{2}{c}{$\omega_\mathrm{0}$ (eV)} & \multicolumn{2}{c}{$\gamma$ (meV)} & \multicolumn{2}{c}{\textit{a}} & \multicolumn{2}{c}{\textit{q}} & \multicolumn{2}{c}{\textit{b}} \\
    \midrule
    Gr/WS$_2$/Gr & 17.5 & \num{2.06611} & 2.05 & \num{14.9059} & 10.0 & \num{0.0134609} & \num{0.003592} & \num{-0.676994} & 1.0 & \num{0.201652} & \num{0.0980037} \\
    Gr/WS$_2$/Gr & 6.1 & \num{2.05627} & 2.05 & \num{10.4881} & \num{10.3497} & \num{0.0856182} & \num{0.0515581} & \num{-0.364795} & \num{-0.752005} & \num{0.187009} & \num{0.174693} \\    
    gr/WS$_2$/gr & 0.69 & \num{2.06617} & 2.065 & \num{23.0983} & \num{18.6268} & \num{0.0514394} & \num{0.0502841} & \num{-0.356203} & \num{-0.325439} & \num{0.111616} & \num{0.115183} \\
    \bottomrule
    \bottomrule
\end{tabularx}
\label{Fano_fit_parameters}
\end{table}

\subsection{Modified Fano-like function fitted to numerical calculations}

\begin{figure}
\centering{\includegraphics[width=\textwidth]{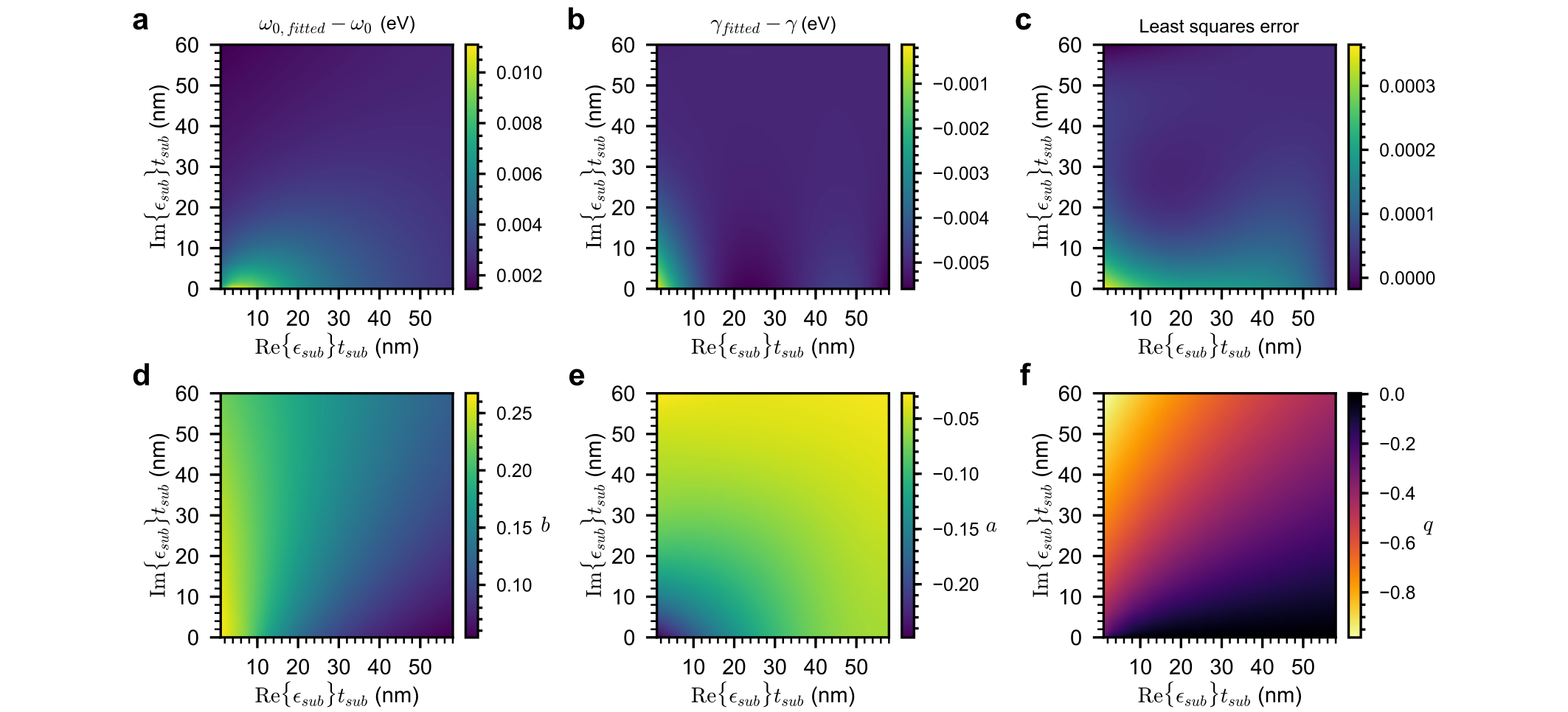}}
\caption{\textbf{Modified Fano-like function fitting of the 2D model:} Evolution of fitted parameters (a) $\omega_\mathrm{fitted}-\omega$, (b) $\gamma_\mathrm{fitted}-\gamma$, (c) the least squares error, (d) \textit{b}, (e) \textit{q}, and (f) \textit{a} of the modified Fano-like function [Eq.~\eqref{Eq:Fano_model}] dependent on the substrate response, specifically by the real versus imaginary part of the dielectric function multiplied by its thickness, $\mathrm{Re}\{\epsilon_\textrm{sub}\}t_\textrm{sub}$ and $\mathrm{Im}\{\epsilon_\textrm{sub}\}t_\textrm{sub}$, respectively.}
\label{Fig:Fano_fit}
\end{figure}

\begin{figure}
\centering{\includegraphics[width=\textwidth]{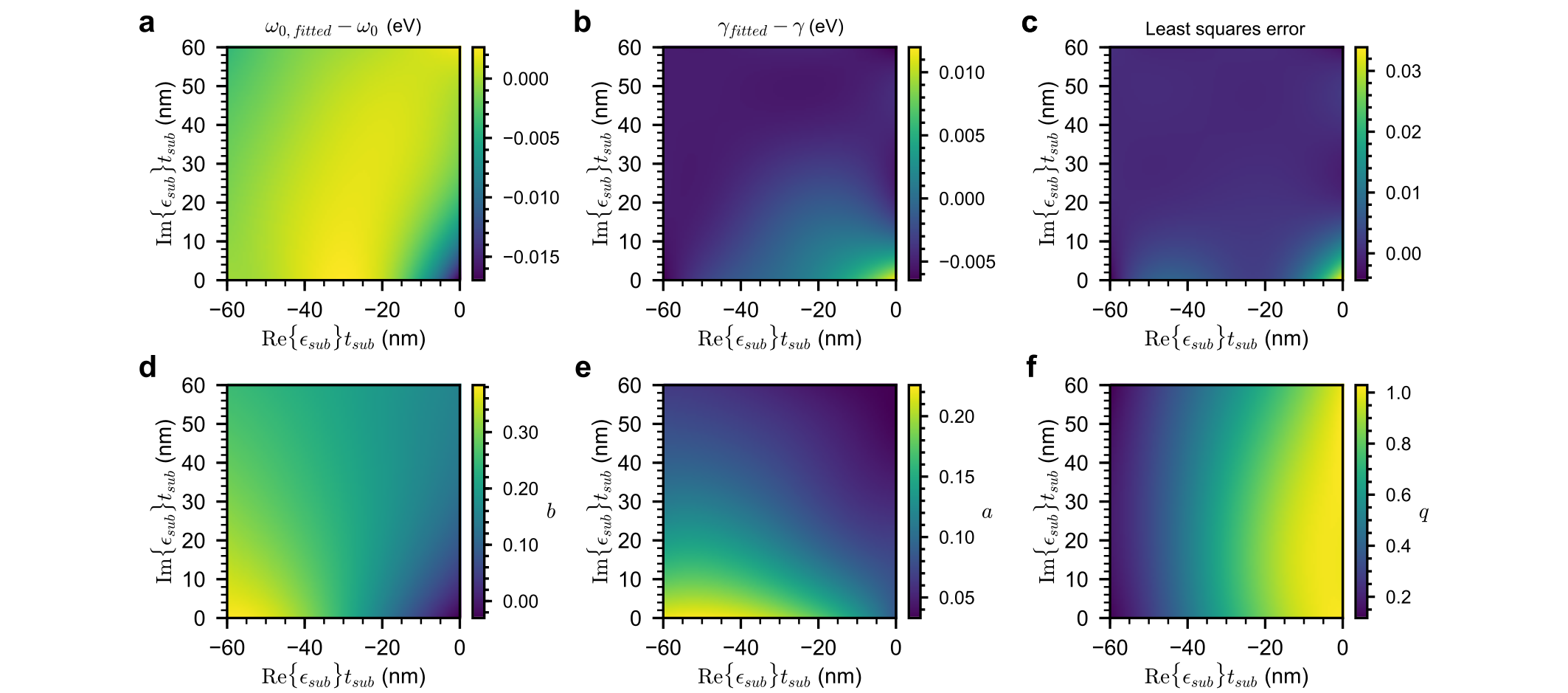}}
\caption{\textbf{Modified Fano-like function fitting of the 2D model for a metallic substrate:} Evolution of fitted parameters (a) $\omega_\mathrm{fitted}-\omega$, (b) $\gamma_\mathrm{fitted}-\gamma$, (c) the least squares error, (d) \textit{b}, (e) \textit{q}, and (f) \textit{a} of the modified Fano-like function [Eq.~\eqref{Eq:Fano_model}] dependent on the substrate response, specifically by the real versus imaginary part of the dielectric function multiplied by its thickness, $\mathrm{Re}\{\epsilon_\textrm{sub}\}t_\textrm{sub}$ and $\mathrm{Im}\{\epsilon_\textrm{sub}\}t_\textrm{sub}$, respectively.}
\label{Fig:Fano_fit_metal}
\end{figure}

To get insight into the behavior of the numerical model [Eq.~\eqref{Eq:Gamma}], the modified Fano-like function [Eq.~\eqref{Eq:Fano_model}] has been fitted to calculations for a dielectric (within its energy band gap,  ${\rm Re}\left\{\epsilon (\omega)\right\} > 0$ and ${\rm Im}\left\{\epsilon (\omega)\right\} > 0$) and a metal (${\rm Re}\left\{\epsilon (\omega)\right\} > 0$ and ${\rm Im}\left\{\epsilon (\omega)\right\} < 0$). The extracted fitted parameters and the least squares error are plotted in Figs. \ref{Fig:Fano_fit}--\ref{Fig:Fano_fit_metal}.  

\end{document}